\documentstyle[12pt,epsfig]{article}

\newlength{\dinwidth}
\newlength{\dinmargin}
\def\be{\begin{equation}}
\def\ee{\end{equation}}
\def\bea{\begin{eqnarray}}
\def\eea{\end{eqnarray}}

\def\lapproxeq{\lower .7ex\hbox{$\;\stackrel{\textstyle
<}{\sim}\;$}}
\def\gapproxeq{\lower .7ex\hbox{$\;\stackrel{\textstyle
>}{\sim}\;$}}

\begin{document}
\titlepage
\begin{flushright}
DTP/96/64  \\
hep-ph/9609448 \\
September 1996 \\
\end{flushright}

\begin{center}
\vspace*{2cm}
{\Large {\bf The QCD description of diffractive $\rho$ meson 
electroproduction}} \\
\vspace*{1cm}
A.\ D.\ Martin, M.\ G.\ Ryskin\footnote{Permanent address:  Laboratory of 
Theoretical Nuclear Physics, St.\ Petersburg Nuclear Physics Institute, 
Gatchina, St.\ Petersburg 188350, Russia.} and T.\ Teubner \\

\vspace*{0.3cm}
Department of Physics, University of Durham, Durham, DH1 3LE, UK. \\

\end{center}

\vspace*{3cm}
\begin{abstract}
We critically review the QCD predictions for the cross sections $\sigma_L$ 
and $\sigma_T$ for diffractive $\rho$ meson electroproduction in 
longitudinally and transversely polarised states in the HERA energy 
region.  We show that both perturbative and non-perturbative
approaches which involve convolution with the $\rho$ meson wave 
function predict values of $\sigma_T$ which fall-off too quickly 
with increasing $Q^2$, in comparison with the data.  We present 
a perturbative QCD model based on the open production of light $q\overline{q}$ 
pairs and parton-hadron duality, which describes all features of the data for 
$\rho$ electroproduction at high $Q^2$ and, in particular, predicts a 
satisfactory $Q^2$ behaviour of $\sigma_L/\sigma_T$.  We find that 
precise measurements of the latter can give valuable information 
on the $Q^2$ behaviour of the gluon distribution at small~$x$. 
\end{abstract}

\newpage

\noindent {\large \bf 1.  Introduction}

The results of the measurements of $\rho$ meson electroproduction, $\gamma^* p 
\rightarrow \rho p$, are intriguing.  These are coming from the H1 \cite{H1} 
and ZEUS \cite{ZEUS,ZEUSW} experiments at the HERA electron-proton collider,
and should be considered in conjunction with the earlier measurements 
of NMC \cite{NMC} at lower energies.  We may briefly summarize the
main features of the observed behaviour of the cross section 
$\sigma (\gamma^* p \rightarrow \rho p)$ as follows:
\begin{tabbing}
(iii)xxx\= $\sigma_L/\sigma_T \sim 2--3$ xxx\= \kill
(i) \> $\sigma \sim 1/Q^5$ \> $\hbox{for} \; 7 < Q^2 < 30$ GeV$^2$. \\
(ii) \> $\sigma \sim W^{0.8}$ \> $\hbox{for} \; 12 < W < 140$ 
GeV.\footnotemark\\
(iii) \> $\sigma_L/\sigma_T \sim 2-4$ \> weakly rising with $Q^2$ for $6 < 
Q^2 < 20$ GeV$^2$. \\
(iv) \> $d\sigma/dt \sim e^{bt}$ \> with $b \simeq 5-6$ GeV$^{-2}$ for
$Q^2 > 10$ GeV$^2$, \\
\> \> as compared to $b \simeq 9$ GeV$^{-2}$ for $Q^2 = 0$.
\end{tabbing}
\footnotetext{This behaviour is observed from the NMC experiment at 
$W \simeq 13$ GeV right through the HERA energy range, 40--140 GeV.}

\noindent As usual, $Q^2$ is the virtuality of the photon, $W$ is the 
centre-of-mass energy of the $\gamma^* p$ system and $t$ is the square
of the four-momentum transfer.  The $\rho$ meson is observed through its
$2\pi$ decay.  
If there are sufficient events, then the angular distribution of the decay 
products allows the measurement of the components $\sigma_L$ and $\sigma_T$ of 
the cross section, which describe $\rho$ production in longitudinally and 
transversely polarised states respectively.  As we shall see, the measurement 
of the $Q^2$ dependence of $\sigma_L/\sigma_T$ is particularly informative.  
The present data, (iii), have large errors, but already indicate the general 
trend.

Observations (ii) and (iv) imply the validity of perturbative QCD for the 
description of high energy $\rho$ electroproduction.  Observation (iv) means 
that the size of the system (the $\gamma^* \rightarrow \rho$ Pomeron vertex) 
decreases with $Q^2$, and that at large $Q^2$ we do indeed have a 
short-distance interaction so that perturbative QCD is justified.  In fact the 
measurement of the slope $b \simeq 5-6$ GeV$^{-2}$ is approximately equal to 
that expected from the size of the proton, which is consistent with the 
hypothesis that at large $Q^2$ the size of the $\gamma^* \rightarrow \rho$ 
vertex is close to zero.  From observation (ii) we see 
that the exponent of the $\sigma \sim W^n$ behaviour has changed from the 
\lq soft' pomeron value $n = 4 (\alpha_P (\overline{t}) - 1) \simeq 
0.2$\footnote{Corresponding to $\alpha_P (0) \simeq 1.08$.} observed 
in $\rho$ photoproduction $(Q^2 = 0)$, to a value $n = 4\lambda \simeq 0.8$ at 
high $Q^2$ which is consistent with the gluon density $xg \sim x^{-\lambda}$ 
extracted\footnote{The MRS parton sets which best describe the recent HERA 
measurements of $F_2$ \cite{F2}, and other data, are 
MRS(A$^\prime$) \cite{MRS} 
and MRS(R2) \cite{MRSR}.  For these the effective value of 
$\lambda$ increases from about 0.2 to 0.3 as $Q^2$ increases from 10 to 50 
GeV$^2$.} from the observed QCD scaling violations of $F_2$.  Moreover
it is in line with the $\sigma \sim W^{0.8}$ behaviour observed in
$J/\psi$ photoproduction, where perturbative QCD is expected to be 
applicable due to the sizeable charm quark mass.

Here we explore the implications of all the observed properties (i)--(iv) for 
the QCD description of $\rho$ electroproduction at HERA.  Before we present 
our detailed study, it is useful to give a brief overview of the situation.  
We begin with the $Q^2$ dependence of $\sigma (\gamma^* p \rightarrow 
\rho p)$.  We will show that for $\rho$ meson electroproduction at high $Q^2$, 
perturbative QCD should be applicable to $\sigma_T$, as well as 
$\sigma_L$.  The leading order perturbative QCD prediction for 
electroproduction in longitudinally polarised states is \cite{RY,BFGMS}
\be
\sigma_L \; \sim \; \frac{[xg (x, Q^2)]^2}{Q^6} \; \sim \; 
\frac{(Q^2)^{2\gamma}}
{Q^6} \; \sim \; \frac{1}{Q^{4.8}}
\label{eq:a1}
\ee
for $Q^2 \gg m_\rho^2$, where $x = Q^2/W^2$ and $\gamma$ is the anomalous 
dimension of the gluon density, $xg (x, Q^2) \sim (Q^2)^\gamma$.  For the 
relevant range of $x$, $10^{-3} \lapproxeq x \lapproxeq 10^{-2}$, we have 
taken\footnote{From the most recent set of partons \cite{MRS,MRSR} we find 
$\gamma \simeq 0.25$ rising to $\gamma \simeq 0.4$ as $x$ decreases from 
$10^{-2}$ to $10^{-3}$ for $Q^2 \approx 10$ GeV$^2$.  Of course in the 
numerical analysis of section 6 the true $x$ and $Q^2$ 
dependence of $\gamma$ is automatically included.}, for the purposes of 
illustration, the representative average value $\gamma = 0.3$.  So the QCD 
prediction for $\sigma_L$ is consistent with the $Q^2$ behaviour 
of the data.  This is not the case for $\sigma_T$.  The prediction for 
$\sigma_T$ appears to be too small and to fall too rapidly with increasing 
$Q^2$.  If only the leading twist component of the light-cone wave 
function\footnote{The twist of the $\rho$ wave function should not be confused 
with that of the operator which corresponds to the $\gamma p$ amplitude.} of 
the $\rho$ is taken into account then
\be
\sigma_T \; \sim \; \frac{m^2}{Q^2} \: \sigma_L \; \sim \; \frac{1}{Q^{6.8}}
\label{eq:a2}
\ee
where $m$ is the current (light) quark mass.  Although the leading twist is 
specified by the QCD sum rules, the next twist is not known.  However, we can 
make reasonable assumptions to estimate its effect.  We find that its 
inclusion has the effect of replacing $m^2$ in (\ref{eq:a2}) by a factor of 
the order of $m_\rho^2$.  Even considering the uncertainties, the value 
predicted for $\sigma_L/\sigma_T$ is still much 
too big and has the wrong $Q^2$ dependence in comparison with the data.  We 
elaborate the above arguments in section 2.

It is frequently claimed that perturbative QCD is not applicable for 
$\sigma_T$ and that its behaviour is of non-perturbative origin, see, 
for example, ref.\ \cite{BFGMS}.  But in this case we 
would expect the same slope $b$ as in photoproduction and a \lq soft' 
$W^{0.2}$ behaviour.  Moreover, non-perturbative QCD predicts a
$1/Q^8$ or stronger 
fall-off of $\sigma_T$ with increasing $Q^2$.  Recall that these features are 
not observed in the data.  A further discussion of the non-perturbative 
approach is given in section 3.

Here we present a resolution of the problem, which is based on the 
application of the hadron-parton duality hypothesis to the production 
of open $q\overline{q}$ pairs.  
First we recall the hadron-parton duality hypothesis for the process $e^+ e^- 
\rightarrow$ hadrons.  In this case the hypothesis gives
\be
\left < \sum_h \: \sigma (e^+ e^- \rightarrow \gamma^* \rightarrow h) \right 
>_{\Delta M^2} \; \simeq \; \left < \sum_q \: \sigma (e^+ e^- \rightarrow 
\gamma^* \rightarrow q\overline{q}) \right >_{\Delta M^2},
\label{eq:a3}
\ee
that is the total hadron production $(h = \rho,\ \omega \ldots)$ averaged over 
a mass interval $\Delta M^2$ (typically $\sim 1$ GeV$^2$) is well 
represented by the partonic cross section.  This duality has been 
checked \cite{PHD} down to 
the lowest values of $\sqrt{s}$.  We may therefore expect the duality to 
apply to diffractive $\rho$ electroproduction for $q\overline{q}$ produced 
in the invariant mass interval containing the $\rho$ meson, $M^2 
\lapproxeq 1-1.5$ GeV$^2$.  In this domain 
the more complicated partonic states $(q\overline{q} + g, \: 
q\overline{q} + 2g, \: q\overline{q} + q\overline{q}, \: \ldots)$ are 
heavily suppressed, while on 
the hadronic side the $2\pi$ (and to a lesser extent the $3\pi$) states are 
known to dominate.  Thus for low $M^2$ we mainly have
\be
\gamma^* \; \rightarrow \; q\overline{q} \; \rightarrow \; 2\pi
\label{eq:a4}
\ee
or in other words
\be
\sigma (\gamma^* p \rightarrow \rho p) \; \simeq \; 0.9 \: \sum_{q = u,d} \: 
\int_{M_a^2}^{M_b^2} \; \frac{d\sigma (\gamma^* p \rightarrow 
(q\overline{q}) p)}{dM^2} \: dM^2
\label{eq:a5}
\ee
where the limits $M_a^2$ and $M_b^2$ are chosen so that they appropriately 
embrace the $\rho$-meson mass region with $M_b^2 - M_a^2 \sim 1$ GeV$^2$.  
The factor 0.9 is included to allow for $\omega$ production.  
This duality model is predictive.  In section 4 we present the QCD formula for 
open $q\overline{q}$ electroproduction via two gluon exchange, and in section 
5 we discuss their general structure.  In particular we show how the scale 
dependence of the gluon density softens the $\sigma_L/\sigma_T \sim Q^2$ 
growth with increasing $Q^2$.  The numerical predictions are presented in 
section 6.  There we calculate diffractive $u\overline{u}$ and $d\overline{d}$ 
electroproduction and use the duality hypothesis to make detailed predictions 
of the $Q^2$ dependence of both $\sigma_L$ and $\sigma_T$ for $\rho$ meson 
electroproduction at HERA; results whose general structure was anticipated in 
the discussion of section 5.

In short, we argue that the convolution of the $q\overline{q}$ wave function 
(produced by the $\gamma^*$) with any reasonable $\rho$ meson wave function 
would yield a prediction for $\sigma_T$ which is in disagreement with
the data.  
Rather we claim that $\rho$ electroproduction proceeds via {\it open} 
$u\overline{u}$, $d\overline{d}$ production at low $M^2$, which has a 
different structure.  Some long time after the interaction with the
proton, confinement distorts the $q\overline{q}$ state and forces it 
to be the $\rho$ meson, as there are no other possibilities.  That is 
the suppression due to the small wave function overlap, $\langle 
q\overline{q}|\rho^0 \rangle$, is not operative.  We depict the 
situation in Fig.~1.

\begin{figure}[htb]
\centerline{\epsfig{file=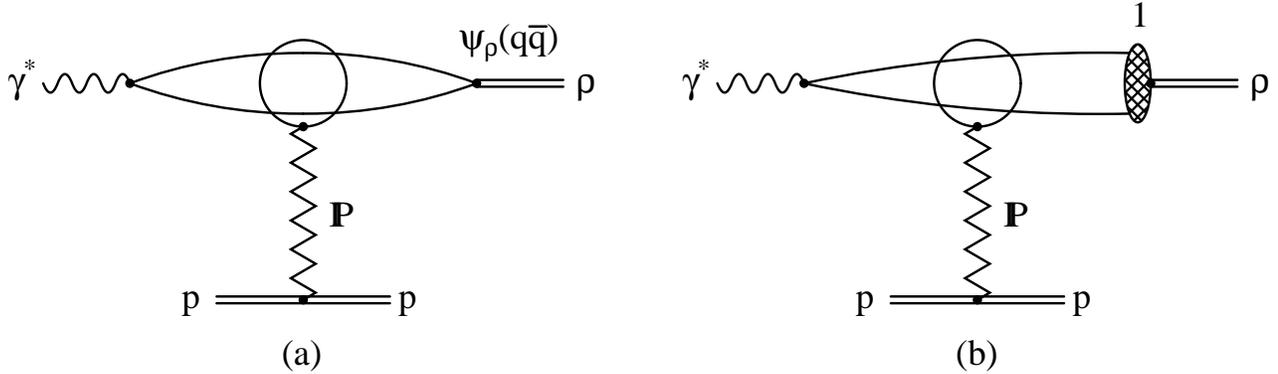,width=17cm}}
\vskip -2mm
\caption[]{ Alternative mechanisms for $\rho$ meson electroproduction:  (a) 
involves a convolution of the $\psi_\gamma (q\overline{q})$ and $\psi_\rho 
(q\overline{q})$ wave functions, whereas (b) is based on open $q\overline{q}$ 
production and parton-hadron duality.  At high $Q^2$ the \lq Pomeron' exchange 
in this picture really stands for the exchange of two gluons in the
$t$ channel.}
\label{fig1}
\end{figure}

\bigskip
\noindent {\large \bf 2.  Standard perturbative approach to the $Q^2$ 
behaviour of $\sigma_{T,L} (\rho)$}

First we wish to sketch the derivation of the perturbative QCD prediction, 
for $\sigma_T$ shown in~(\ref{eq:a2}),
\be
\sigma_T (\gamma^* p \rightarrow \rho p) \; \sim \; [xg (x, Q^2)]^2/Q^8
\label{eq:a6}
\ee
for $Q^2 \gg m_\rho^2$, and to show that it is infrared stable.  We must 
therefore study the $\gamma^* \rightarrow \rho$ Pomeron vertex (or so-called 
impact factor) of Fig.~1(a), which we denote by $J_T$.  We shall also consider 
$J_L$.  The factors are given by the convolution of the wave functions 
$\psi_\gamma (q\overline{q})$ and $\psi_\rho (q\overline{q})$.  It is 
found that \cite{BFGMS,IV}
\be
J_i \; = \; f_\rho \: \int \: \frac{dz \: dk_T^2}{\varepsilon^2 + k_T^2} \: 
\psi_\rho^i (z, k_T^2) B_i
\label{eq:a7}
\ee
with $i = T$ or $L$.  The quantity $f_\rho$ is the $\rho$ meson decay constant 
and the term $\varepsilon^2$ in the quark propagator is
\be
\varepsilon^2 \; = \; z (1 - z) Q^2 \: + \: m^2
\label{eq:a8}
\ee
where $m$ is the current quark mass.  $B_i$ 
are the helicity factors\footnote{The vertex satisfies $s$ channel (quark) 
helicity conservation.  In general for $t \neq 0$ we would also have 
off-diagonal helicity factors, $B (\gamma_T, \rho_L)$ and 
$B (\gamma_L, \rho_T)$.} coming from the quark loop, see Fig.~1(a),
\bea
\label{eq:a9}
B_L & = & 2z (1 - z) \: \sqrt{Q^2}, \\
\label{eq:a10}
B_T & = & -m.
\eea
$\psi_\rho (z, k_T^2)$ is the momentum representation of the $\rho$ meson wave 
function; $z$ and $\mbox{\boldmath $k$}_T$ are the Sudakov and transverse 
momentum components carried by one of the quarks with respect to the photon.  
The other quark has components $1-z$ and $-\mbox{\boldmath $k$}_T$.

The wave functions $\psi_\rho^{L,T}$ decrease slowly with $k_T^2$ and the 
convergence of the integral in (\ref{eq:a7}) is provided only by the 
denominator $\varepsilon^2 + k_T^2$.  We therefore introduce an 
integrated wave function
\be
\phi_\rho^i (z) \; \equiv \; \int^{\varepsilon^2} \: dk_T^2 \: \psi_\rho^i 
(z, k_T^2)
\label{eq:a11}
\ee
defined by the scale $\mu^2 = \varepsilon^2$ at which the integral ceases to 
converge.  The quantities $\phi_\rho^i$ are called the leading twist 
light-cone 
$\rho$ meson wave functions and have been well studied in the framework of QCD 
sum rules \cite{CZ,BB}.  As $Q^2 \rightarrow \infty$ (that is $\varepsilon^2 
\rightarrow \infty$) we have
\be
\phi_\rho^i (z) \; \rightarrow \; 6z (1 - z)
\label{eq:a12}
\ee
for both $i = T,\ L$.  Their behaviour at finite scales can be found in Refs.\ 
\cite{CZ,BB}, but in any 
case the $\phi_\rho$ vanish at least as fast as $z$ as $z \rightarrow 0$ and 
$1 - z$ as $z \rightarrow 1$.  We may rewrite the impact factors (\ref{eq:a7}) 
in terms of the integrated wave functions $\phi_\rho^i (z)$.  We obtain
\be
J_i \; \approx \; f_\rho \: \int \: \frac{dz}{\varepsilon^2} \:
\phi_\rho^i (z) B_i.
\label{eq:a13}
\ee
Finally, we must convolute $J_i$ with the $q\overline{q}$-proton interaction 
amplitude $T$ given by the BFKL Pomeron (or two-gluon exchange ladder).  The 
amplitude $T$ behaves as
\be
\frac{1}{s} {\rm Im} T \; = \; \sigma_{q\overline{q} - p} \; \sim \; 
\frac{xg (x, \varepsilon^2)}{\varepsilon^2} \; \sim \; (\varepsilon^2)^{\gamma 
- 1}
\label{eq:a14}
\ee
where recall that the scale is $\varepsilon^2 = z (1 - z) Q^2 + m^2$, 
and where $\gamma (x)$ is the anomalous dimension of the gluon.  Thus the 
amplitudes for $\rho$ electroproduction from transversely $(i = T)$ and 
longitudinally $(i = L)$ polarised photons are 
\be
A_i \; = \; J_i \: \otimes \: T \; = \; f_\rho \: \int \: 
\frac{dz}{(\varepsilon^2)^{2 - \gamma}} \: \phi_\rho^i (z) \: B_i,
\label{eq:a15}
\ee
which yield the following $Q^2$ behaviour of the cross sections
\bea
\label{eq:a16}
\sigma_T \; \sim \; |A_T|^2 & \sim & (m/(Q^2)^{2 - \gamma})^2 \; \sim \; 
m^2/Q^{6.8}, \\
\label{eq:a17}
\sigma_L \; \sim \; |A_L|^2 & \sim & Q^2 (1/(Q^2)^{2 - \gamma})^2 \; \sim \; 
1/Q^{4.8}.
\eea
For illustration, we have again set the gluon anomalous dimension $\gamma 
= 0.3$.  We emphasize that the integral in (\ref{eq:a15}) is
convergent for $A_T$ (for any $\gamma > 0$), as well as 
for $A_L$.  Thus $\varepsilon^2 \sim Q^2$ and 
perturbative QCD is valid not only for $\sigma_L$ (where we have additional 
convergence due to $B_L \sim z (1 - z)$), but {\it also} for $\sigma_T$.

We note that while the prediction for the relative $Q^2$ dependence 
of $\sigma_T$ and $\sigma_L$ is meaningful (although not supported 
by the data), the value for the ratio
\be
\frac{\sigma_T}{\sigma_L} \; \sim \; \frac{m^2}{Q^2}
\label{eq:b17}
\ee
(which is in gross disagreement with the data) is not a reliable 
estimate.  The 
reason is that the current $u,d$ quark masses are very small ($m \lapproxeq 7$ 
MeV) and that therefore we must consider how the non-leading twist 
contribution to $\psi_\rho^T (z, k_T^2)$ will modify the prediction 
for $\sigma_T$.  The non-leading twist is not known.  However, it is 
reasonable to assume that instead of two variables, the $\rho$ wave 
function $\psi_\rho^T$ depends on only one variable, namely the 
invariant mass of the $q\overline{q}$ pair\footnote{This hypothesis 
is very natural from a dispersion relation viewpoint \cite{GFS}.}
\be
M^2 \; = \; \frac{k_T^2}{z (1 - z)},
\label{eq:a18}
\ee
where we neglect $m^2$.  Then, after some algebra, it is possible to show that 
the impact factor $J_T$ can be written in the form of (\ref{eq:a15}) with 
$\phi_\rho^T = 6z (1 - z)$, and that the helicity factor becomes
\be
B_T \; = \; - {\textstyle \frac{1}{2}} \: m_\rho \: \left [ z^2 + (1 - z)^2 
\right ],
\label{eq:a19}
\ee
rather than the very small \lq leading-twist' prediction given 
in (\ref{eq:a10}).  The reason that we still obtain a definite
prediction for $J_T$, again in terms of $f_\rho$, is due 
to the fact that this same non-leading twist component of $\psi_\rho^T$ 
describes the decay $\rho_T \rightarrow e^+ e^-$, that is the $k_T$ integral 
over the quark loop describing the $\rho_T$ decay is the same integral
that occurs in the impact factor $J_T$ for $Q^2 \gg m_\rho^2$.  In
this way we are able to normalise the non-leading twist to the
observed width of the decay, that is to the decay constant $f_\rho$.

If we estimate the $\rho$ electroproduction amplitude $A_T$ of (\ref{eq:a15}) 
using the modified form (\ref{eq:a19}) of $B_T$ then we obtain
\be
\frac{\sigma^T}{\sigma_L} \; = \; c \: \frac{m_\rho^2}{Q^2}
\label{eq:a20}
\ee
with $c \sim 2$.  The precise value of $c$ depends on the actual forms of 
$\phi_\rho^{T,L} (z)$ at the experimentally relevant scales, $\mu^2 \sim 10$ 
GeV$^2$, which are far from the asymptotic region where $\phi_\rho^{T,L} (z) = 
6z (1 - z)$.  In our approximate estimate of $c \sim 2$ we have used 
the $\phi_\rho^{T,L} (z)$ wave functions of ref.\ \cite{BB}.  Although a 
considerable improvement on (\ref{eq:b17}), the prediction (\ref{eq:a20}) for 
the ratio $\sigma^T/\sigma^L$ is still much smaller than the observed 
ratio, and as before decreases more rapidly with $Q^2$ than indicated 
by the data \cite{H1,ZEUS,NMC}.  In short, the standard perturbative 
QCD predictions for $\sigma_T (\gamma^* p \rightarrow \rho p)$ are not
in agreement with the observations. \\

\noindent {\large \bf 3.  Non-perturbative approach to the $Q^2$ dependence of 
$\sigma_T (\rho)$}

It has been argued that the main contribution to $\sigma_T$ comes from the 
non-perturbative region~\cite{BFGMS}.  Let us disregard the fact that
the perturbative integral (\ref{eq:a15}) is convergent for $\sigma_T$
and suppose that non-perturbative effects dominate.  In order to 
obtain non-perturbative contributions associated with 
small $(\sim \mu^2)$ virtualities, we must get contributions from 
the end-point regions of integration
\be
z \; \lapproxeq \; \mu^2/Q^2 \quad\quad \hbox{and} \quad\quad 1 - z \;
\lapproxeq \; \mu^2/Q^2.
\label{eq:a21}
\ee
Only then will we sample small scales $\varepsilon^2 \sim \mu^2$ and large 
distances $\rho \sim 1/\varepsilon \sim 1/\mu$.  However, for large distances 
the quark effectively has a constituent mass $m_q \sim \frac{1}{2} m_\rho$ and 
the non-relativistic wave function, $\phi_\rho^T (z) \sim \delta (z - 
\frac{1}{2})$, is appropriate.  Certainly $\phi_\rho^T (z)$ decreases 
exponentially, or at least as a large power, as $z \rightarrow 0$ or 
$z \rightarrow 1$.  Thus the contribution from the 
regions (\ref{eq:a21}) should be strongly 
suppressed.  Even if $\phi_\rho^T (z) \sim z (1 - z)$, as in (\ref{eq:a12}), 
we would obtain from (\ref{eq:a13}) with $\varepsilon \sim \mu$
\be
\sigma_T (\hbox{non-pert.}) \; \sim \; \left [ \frac{1}{\mu^2} \: 
\int_0^{\mu^2/Q^2} \: dz \: \phi_\rho^T (z) \: B_T \right ]^2 \; \sim \; 
\frac{1}{Q^8}.
\label{eq:a22}
\ee
Thus for the actual non-perturbative prediction we would expect an even faster 
fall-off with increasing $Q^2$. \\

\noindent {\large \bf 4.  QCD model for $\sigma_{L,T} (\rho)$ via open 
$q\overline{q}$ production}

The above discussion suggests that the problem in successfully 
describing $\rho$ meson electroproduction may be associated with 
having to convolute with a $\rho$ 
meson wave function, which inevitably leads to a form-factor-like suppression 
of the form $\left | \langle q\overline{q}| \rho^0 \rangle \right |^2 \sim 
1/Q^4$.  Here we study an alternative and physically compelling mechanism for 
$\rho$ electroproduction based on the production of $u\overline{u}$ and 
$d\overline{d}$ pairs in a broad mass interval containing the $\rho$ 
meson.  In 
this mass interval phase space forces these $q\overline{q}$ pairs to hadronize 
dominantly into $2\pi$ states, with only a small amount of $3 \pi$ 
production.  
Moreover, provided the $q\overline{q}$-proton interaction does not distort the 
spin, we expect that the process $\gamma^* \rightarrow q\overline{q} 
\rightarrow 2 \pi$ will dominantly produce $2\pi$ systems with $J^P = 1^-$.  
The calculation of the diffractive electroproduction of $q\overline{q}$ pairs 
therefore allows, via the parton-hadron duality hypothesis, a detailed 
prediction of the structure of $\rho$ meson electroproduction.

\begin{figure}[htb]
\centerline{\epsfig{file=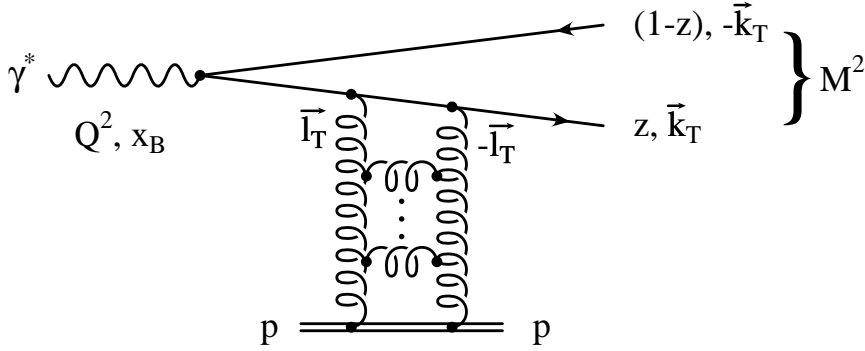,width=12cm}}
\caption[]{ Diffractive open $q\overline{q}$ production in 
high energy $\gamma^* 
p$ collisions, where $z$ is the fraction of the energy of the photon that is 
carried by the quark.  The transverse momenta of the outgoing quarks are 
$\pm \vec{k}_T$, and those of the exchanged gluons are 
$\pm \vec{\ell}_T$.}
\label{fig2}
\end{figure}

The formula for the diffractive production of open $q\overline{q}$ pairs is 
given in ref.\ \cite{LMRT,NN}.  For light quarks we may safely put the
current quark 
mass $m = 0$.  The process is shown in Fig.~2.  We use the same notation as in 
ref.\ \cite{LMRT}, so the scale at which the gluon distribution is sampled is 
denoted
\be
K^2 \; = \; z (1 - z) Q^2 \: + \: k_T^2 \; = \; \frac{k_T^2}{1 - \beta},
\label{eq:a24}
\ee
where the last equality follows since $z (1 - z) = k_T^2/M^2$ and
\be
\beta \; \equiv \; \frac{Q^2}{Q^2 + M^2}.
\label{eq:a25}
\ee
Note that the scale $K^2$ plays the role that $\varepsilon^2$ played 
for exclusive vector 
meson production (cf.\ (\ref{eq:a8})), and that it determines the transverse 
distances $b_T \sim 1/K$ that are typically sampled in the process.  It is 
convenient to replace the $dk_T^2$ integration over the quark
transverse momenta $k_T$ in formulae (40, 41) in ref.\ \cite{LMRT} by
an integration over $dK^2$.  
Then it is straightforward to show that these formulae giving the 
$\gamma_{L,T}^* p \rightarrow (q\overline{q}) p$ cross sections in the forward 
direction $(t = 0)$ may be written in the form
\bea
\label{eq:a27}
\frac{d^2 \sigma_L}{dM^2 dt} & = & \frac{4 \pi^2 e_q^2 \alpha}{3} \: 
\frac{Q^2}{(Q^2 + M^2)^4} \: \int_{K_0^2}^{\frac{1}{4} (Q^2 + M^2)} \: 
\frac{dK^2 \: K^2}{\sqrt{1 - 4K^2/(Q^2 + M^2)}} \; [I_L (K^2) ]^2, \\
\label{eq:a28}
\frac{d^2 \sigma_T}{dM^2 dt} & = & \frac{4 \pi^2 e_q^2 \alpha}{3} \: 
\frac{M^2}{(Q^2 + M^2)^3} \: \int_{K_0^2}^{\frac{1}{4} (Q^2 + M^2)} \: 
\frac{dK^2 (1 - 2\beta K^2/Q^2)}{\sqrt{1 - 4K^2/(Q^2 + M^2)}} \; [I_T (K^2) ]^2
\eea
where $\alpha$ is the electromagnetic coupling.  The quantities 
$I_{L,T}$ are the integrations over the transverse momenta, $\pm 
\mbox{\boldmath $\ell$}_T$, of the exchanged gluons (see Fig.~2)
\bea
\label{eq:a29}
I_L (K^2) & = & K^2 \: \int \: \frac{d\ell_T^2}{\ell_T^4} \: \alpha_S 
(\ell_T^2) \: f (x, \ell_T^2) \: \left ( \frac{1}{K^2} \: - \: 
\frac{1}{K_\ell^2} \right ), \\
\label{eq:a30}
I_T (K^2) & = & \frac{K^2}{2} \: \int \: \frac{d\ell_T^2}{\ell_T^4} \:
\alpha_S (\ell_T^2) \: f (x, \ell_T^2) \: \left ( \frac{1}{K^2} \: -
\: \frac{1}{2k_T^2} \: + \: \frac{K^2 - 2k_T^2 + \ell_T^2}{2k_T^2 \: 
K_\ell^2} \right ),
\eea
where $x = (Q^2 + M^2)/W^2$,
\be
K_\ell^2 \; \equiv \; \sqrt{(K^2 + \ell_T^2)^2 \: - \: 4k_T^2 \ell_T^2},
\ee
and $f (x, \ell_T^2)$ is the unintegrated gluon distribution of the 
proton.  We 
will use formulae (\ref{eq:a27}) and (\ref{eq:a28}) to predict $\rho$ meson 
electroproduction.  They involve integration over the quark $k_T^2$ (or $K^2$) 
and over the $\ell_T^2$ of the exchanged gluons.  As we are dealing with a 
diffractive process we see that the cross sections have a quadratic 
sensitivity to the gluon density.

It is useful to inspect the leading $\ln K^2$ approximation to the $d\ell_T^2$ 
integrations of (\ref{eq:a29}) and (\ref{eq:a30}).  In this
approximation it is 
assumed that the main contributions to the integrals come from the domain 
$\ell_T^2 \lapproxeq K^2$, and so on expanding the integrands we obtain
\be
I_L^{LLA} \; = \; I_T^{LLA} \; = \; \frac{\alpha_S (K^2)}{K^2} \: 
\int^{K^2} \: 
\frac{d\ell_T^2}{\ell_T^2} \: f (x, \ell_T^2) \; = \; \frac{\alpha_S (K^2)}
{K^2} \: xg (x, K^2).
\label{eq:a31}
\ee
By analogy with (\ref{eq:a14}), we see that $I_{L,T}$ are essentially the 
cross sections for the $(q\overline{q})_{L,T}$ interaction with the proton.  
Of course, in the calculations presented in section 6 we do not use the 
leading $\log$ approximation, but instead we perform the explicit 
$d\ell_T^2$ integrations over $f(x, \ell_T^2) = \partial 
(xg (x, \ell_T^2))/\partial \ln \ell_T^2$ given in (\ref{eq:a29}) and 
(\ref{eq:a30}).  We treat the infrared region using 
the linear approximation described in ref.\ \cite{LMRT} for 
low $\ell_T^2$ values (that is $\ell_T^2 < \ell_0^2$).  We find 
stability of the results to reasonable variations of the choice 
of $\ell_0^2$. \\

\noindent {\large \bf 5.  Insight into the structure of the cross sections 
$\sigma_{L,T}$}

In section 6 we show the predictions for the $Q^2$ behaviour of $\sigma_L$ 
and $\sigma_T$ for $\rho$ electroproduction, which are obtained from the 
numerical evaluation of (\ref{eq:a27}) and (\ref{eq:a28}) integrated over the 
$\rho$ mass region.  However, it is informative to anticipate some of the 
general features of the results.  First we study the infrared convergence of 
the $dK^2$ integrations of (\ref{eq:a27}) and (\ref{eq:a28}).  We note from 
the approximate forms of $I_{L,T}$ in (\ref{eq:a31}) that
\be
I_i \; \sim \; x^{- \lambda} \: (K^2)^\gamma/K^2
\label{eq:a32}
\ee
where $\lambda$ and $\gamma$ are the effective exponents of the gluon 
defined by
\be
xg (x, K^2) \; \sim \; x^{- \lambda} \: (K^2)^\gamma.
\label{eq:a33}
\ee
\begin{figure}[htb]
\centerline{\epsfig{file=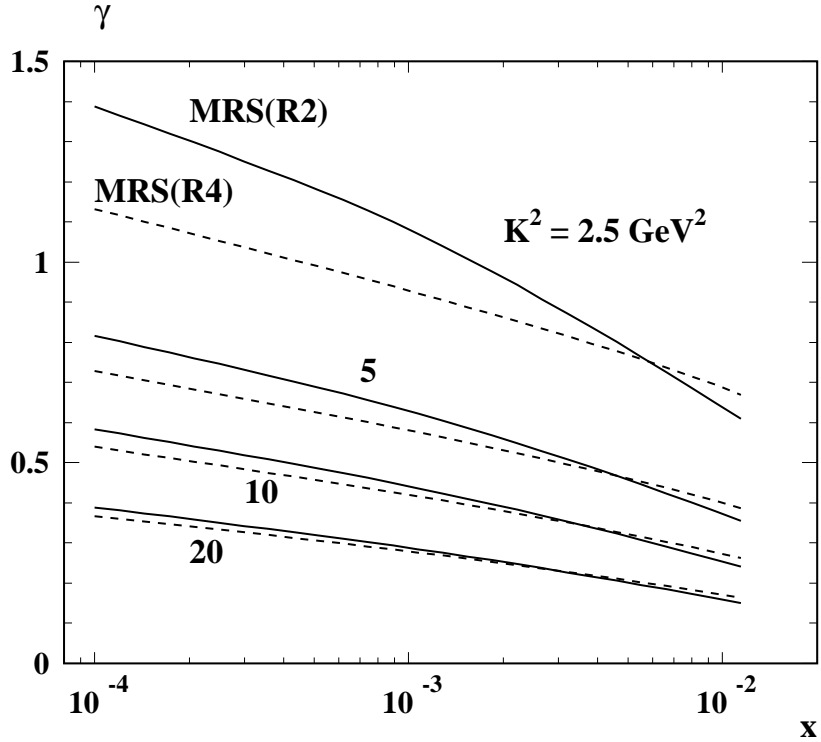,width=12cm}}
\vskip -7mm
\caption[]{ The continuous curves show the effective anomalous 
dimension $\gamma$ 
of the gluon (defined by $xg (x, K^2) \sim (K^2)^\gamma$) determined from the 
MRS(R2) set of partons \protect{\cite{MRSR}} for $K^2 = 2.5, 5, 10$ and 20 
GeV$^2$.  The dashed curves correspond to the values of $\gamma$ for the R4 
set of partons.}
\label{fig3}
\end{figure}
We see that the integration (\ref{eq:a27}) is infrared convergent 
provided that 
$\gamma > 0$ as $K^2 \rightarrow 0$, whereas we require $\gamma > 0.5$
to ensure 
the convergence of (\ref{eq:a28}).  How does the 
value of $\gamma$ depend on $K^2$?  At high energy $W$ (that is 
$x \approx Q^2/W^2 
\rightarrow 0$) the gluon $g (x, K^2)$ increases much faster as $x$ decreases 
for large $K^2$ ($xg \sim x^{- \lambda}$ with $\lambda \gapproxeq 0.3$) than 
for small $K^2$.  Thus the effective anomalous dimension $\gamma$ increases 
when $x$ and $K^2$ decrease.  The behaviour is evident in Fig.~3 which shows 
the values of $\gamma$ (as a function of $x$ for selected $K^2$) obtained 
from two recent sets of partons.  For example, let us take a typical value 
$x \approx Q^2/W^2 = 10^{-3}$ relevant for the measurements at HERA 
(say $Q^2 = 10$ GeV$^2$ and $W = 100$ GeV).  We see from Fig.~3 that 
$\gamma$ increases from 0.3, 0.45, 0.6 to 1 as $K^2$ decreases from 
about 20, 10, 5 to 2.5 GeV$^2$.  The infrared 
convergence requirement, $\gamma > 0.5$, of (\ref{eq:a28}) is therefore 
already satisfied when $K^2$ has decreased to 8 GeV$^2$.  In general, the 
behaviour of $\gamma$ with $K^2$ 
amply provides, via (\ref{eq:a32}), the infrared convergence 
of (\ref{eq:a28}), 
as well as of (\ref{eq:a27}).  This explains the reason why our numerical 
evaluation of (\ref{eq:a28}) for $\sigma_T$ depends only weakly on the
infrared cut-off\footnote{In section 6 we choose $K_0 = 0.2$ GeV, the 
order of the inverse confinement radius 1 fm$^{-1}$.} $K_0^2$.  
Indeed integral (\ref{eq:a27}) for $\sigma_L$ is controlled by contributions 
close to the upper limit and we expect
\be
\frac{d^2 \sigma_L}{dM^2 dt} \; \sim \; \frac{(Q^2)^{2\gamma - 2\lambda}}{Q^6}.
\label{eq:a34}
\ee
This is exactly the same $Q^2$ behaviour as the prediction (\ref{eq:a17}) for 
{\it exclusive} $\rho_L$ electroproduction; here we have been more precise and 
displayed the $Q^2$ dependence coming from the $x (\simeq Q^2/W^2)$ behaviour 
of the gluon.  Of course, the result (\ref{eq:a34}) 
is very approximate and the detailed dependence of the $Q^2$ behaviour of 
$\sigma_L$ (as well as $\sigma_T$) on the properties of the gluon must await 
the numerical predictions of section 6.

Nevertheless we can take the general discussion further and anticipate the 
main features of the $Q^2$ 
behaviour of the important ratio $\sigma_L/\sigma_T$.  We first rewrite 
(\ref{eq:a27}) and (\ref{eq:a28}) in terms of an integration over 
the angles of 
the produced $q\overline{q}$ pair.  We use the polar angle $\theta$ of the 
outgoing $q$ in $q\overline{q}$ rest frame with respect to the incident 
direction of the proton.  Thus we have
\be
k_T \; = \; \textstyle{\frac{1}{2}} M \sin \theta,
\label{eq:a35}
\ee
and the square root in the 
denominators of (\ref{eq:a27}) and (\ref{eq:a28}) is 
equal to $\cos \theta$.  Also the factor in the numerator of (\ref{eq:a28})
\be
1 \: - \: 2\beta K^2/Q^2 \; = \; \textstyle{\frac{1}{2}} \: 
(1 + \cos^2 \theta) 
\; = \; |d_{11}^1 (\theta)|^2 \: + \: |d_{1-1}^1 (\theta)|^2,
\label{eq:36}
\ee
where $d_{\lambda \mu}^J (\theta)$ are the conventional spin 
rotation matrices.  Then equations (\ref{eq:a27}) and (\ref{eq:a28}) become
\bea
\label{eq:a37}
\frac{d^2 \sigma_L}{dM^2 dt} & = & \frac{4 \pi^2 e_q^2 \alpha}{3} \: 
\frac{Q^2}{(Q^2 + M^2)^2} \: \frac{1}{8} \: \int_{-1}^1 \: d \cos \theta \: 
\left | d_{10}^1 (\theta) \right |^2 \: \left | I_L \right |^2, \\
& & \nonumber\\
\label{eq:a38}
\frac{d^2 \sigma_T}{dM^2 dt} & = & \frac{4 \pi^2 e_q^2 \alpha}{3} \: 
\frac{M^2}{(Q^2 + M^2)^2} \: \frac{1}{4} \: \int_{-1}^1 \: d \cos \theta \: 
\left ( \left | d_{11}^1 (\theta) \right |^2 \: + \: \left |d_{1-1}^1 
(\theta) \right |^2 \right ) \: \left | I_T \right |^2,
\eea
where the dependence on the rotation matrices appropriately reflects the decay 
of the $\rho$ meson from longitudinally and transversely polarised states, 
respectively.

In the limit of no interaction with the proton (that is $I_L = I_T =$ 
constant) the photon has to produce the $q\overline{q}$ pair in a pure
spin $J = 1$ state.  We immediately find from (\ref{eq:a37}) and 
(\ref{eq:a38}) that
\be
\frac{\sigma_L}{\sigma_T} \; = \; \frac{Q^2}{2M^2} \: \frac{\int d \cos \theta 
\: \sin^2 \theta}{\int d \cos \theta \: (1 + \cos^2 \theta)} \; = \; 
\frac{Q^2}{4M^2}.
\label{eq:a39}
\ee
In the realistic situation, the two-gluon exchange interaction distorts the 
$q\overline{q}$ state produced by the \lq heavy' photon.  Some idea of the 
consequences of this distortion can be anticipated from the leading $\log$ 
approximation (\ref{eq:a31}) for $I_L$ and $I_T$, in which
\be
I_L \; = \; I_T \; \sim \; \frac{(K^2)^\gamma}{K^2} \; \sim \; \frac{1}
{(\sin^2 \theta)^{1 - \gamma}}.
\label{eq:a40}
\ee
We substitute this behaviour into (\ref{eq:a37}) and (\ref{eq:a38}), and 
project\footnote{To be precise the rotation matrices $D_{\lambda \mu}^J 
(\phi, \theta, -\phi)$ form the orthogonal basis and we project out the 
components $c (\lambda)$ from the $q\overline{q}$ amplitudes $D_{1 
\lambda}^{1 *} I (\theta)$ with the matrix $D_{1 \lambda}^1$.  However, the 
$\phi$ integrations are trivial and hence the projection can be done simply 
in terms of $d_{1 \lambda}^1$.} out the spin 1 components of the underlying 
$q\overline{q}$ production amplitudes ($\sim d_{1 \lambda}^1 (\theta) \: 
I (\theta)$ where $I_L = I_T \equiv I (\theta) \sim (\sin^2 
\theta)^{\gamma - 1})$.  We then use the identity
\be
\int_0^\pi \: \sin^p \theta \: d\theta \; = \; \sqrt{\pi} \: \frac{\Gamma (
\frac{1}{2} + \frac{1}{2}p)}{\Gamma (1 + \frac{1}{2}p)}
\label{eq:a41}
\ee
to evaluate the projections
\be
c (\lambda) \; = \; \frac{2J + 1}{2} \: \int \: d \cos \theta \: \left (d_{1 
\lambda}^{J = 1} (\theta) \: I (\theta) \right ) \: d_{1 \lambda}^{J = 1} 
(\theta),
\label{eq:a42}
\ee
assuming that $\gamma$ is a constant over the region of integration.  With 
this assumption we find the interesting result
\be
\frac{\sigma_L}{\sigma_T} \; = \; \frac{Q^2}{2M^2} \: \frac{|c 
(\lambda = 0)|^2}
{|c (\lambda = 1)|^2 \: + \: |c (\lambda = -1)|^2} \; = \; \frac{Q^2}{M^2} \: 
\left (\frac{\gamma}{\gamma + 1} \right )^2.
\label{eq:a43}
\ee
The dependence on $\gamma$ has the effect of masking the $Q^2$ growth of 
$\sigma_L/\sigma_T$.  This can be seen by inspecting Fig.~3 -- higher $Q^2$ 
means larger $x$ and both changes imply smaller $\gamma$.  The 
projection integrals (\ref{eq:a42}) for the amplitudes (with 
their linear dependence on $I_i (\theta)$) 
are more infrared convergent than (\ref{eq:a37}) and (\ref{eq:a38}).  Now 
$\sigma_T$ (as well as $\sigma_L$) is convergent provided only 
that $\gamma > 0$ 
as $K^2 \rightarrow 0$ (that is as $\theta \rightarrow 0$).  In fact, provided 
$x$ remains 
sufficiently small, both the $\sigma_L$ and $\sigma_T$ integrations receive 
their main contributions from the region $K^2 \lapproxeq Q^2/4$, and so we 
should insert into (\ref{eq:a43}) the average $\gamma$ sampled in this $x$, 
$K^2$ domain.  Indeed the decrease of $\gamma$ with increasing $K^2 \lapproxeq 
Q^2/4$ is found to considerably suppress the growth of
$\sigma_L/\sigma_T$ with 
increasing $Q^2$, and to largely remove the gross disagreement of the QCD 
prediction with the data; see the full numerical calculation presented in 
section 6.  We may turn the argument the other 
way round.  Accurate measurements 
of the ratio $\sigma_L/\sigma_T$ as a function of $Q^2$ will offer 
an excellent 
way of constraining the $K^2$ and $x$ behaviour of the gluon $g (x, K^2)$ 
in the region $K^2 \lapproxeq Q^2/4$ and $x \approx Q^2/W^2$.  Of 
course result (\ref{eq:a43}), which is based on a constant $\gamma$,
is oversimplified.  It is given only to indicate the general trend.
The full calculation of section 6 
is performed with a realistic gluon distribution and so automatically 
allows for the $K^2$ (and $x$) dependence of $\gamma$.

We see that the projection integrals (\ref{eq:a42}) converge in the infrared 
region of small $K^2 \approx \frac{1}{4} Q^2 \sin^2 \theta$ (that is at small 
$\theta$) for any $\gamma > 0$, even for $\sigma_T$ (that is for $c (\lambda = 
\pm 1)$).  We have stronger infrared convergence for $\sigma_L$ or $c (\lambda 
= 0)$ due to $d_{10}^1 = - \sin \theta/\sqrt{2}$.  We also notice that the 
factor $I (\theta) = 1/(\sin^2 \theta)^{1 - \gamma}$, arising from the 
$q\overline{q}$-proton interaction, gives a strong peak in the forward 
direction\footnote{The height of the peak is limited by the infrared cut-off, 
$K_0 = 0.2$ GeV, provided by confinement.}.  It means that the 
distortion caused by the interaction will, in principle, produce higher spin 
$q\overline{q}$ states.  Most probably the higher spin states at small $M^2$ 
are killed by confinement during the hadronization stage as there is 
insufficient phase space to create $2\pi$ states with large spin with $M^2 
\lapproxeq 1$ GeV$^2$.  In any case the higher spin components\footnote{Indeed 
it will be interesting to study 
the detailed spin decomposition of $\gamma^* \rightarrow$ open $q\overline{q}$ 
production as a function of $M^2$.  In this way we can investigate how the QCD 
\lq Pomeron' distorts the initial state and how confinement/parton-hadron 
duality operates in different (relatively small) $M^2$ regions for 
the different 
$J^P$ states.} cannot affect 
$\rho$ production, since confinement cannot change the spin of the produced 
$q\overline{q}$ state.  At higher energies (small $x$) the anomalous dimension 
$\gamma$ grows and the function $I (\theta)$ is not so singular as $\theta 
\rightarrow 0$.  Therefore in this energy domain the incoming spin of the 
$q\overline{q}$ system is not so contaminated by $J \neq 1$ 
components arising from the interaction with the proton.  In the black disk 
limit of the proton when the cross section approaches the 
saturation (unitarity) 
limit $\gamma$ tends to 1 and we come back to pure $J = 1$ $q\overline{q}$ 
production.

\noindent {\large \bf 6.  Numerical QCD predictions for $\rho$ 
electroproduction}

We use parton-hadron duality to predict $\rho$ electroproduction from the QCD 
formulae for open $u\overline{u}$ and $d\overline{d}$ production.  To 
be precise we compute
\be
\sigma_{L,T} (\rho) \; = \; 0.9 \: 
\int_{(0.6\ {\rm GeV})^2}^{(1.05\ {\rm GeV})^2}
\: dM^2 \: \frac{d\sigma_{L,T} (J = 1)}{dM^2}
\label{eq:a44}
\ee
where $d\sigma_{L,T} (J = 1)/dM^2$ are the spin 1 projections of 
open $q\overline{q}$ 
production of (\ref{eq:a37}) and (\ref{eq:a38}), carried out as described in 
(\ref{eq:a42}), and where the cross sections have been integrated over $t$ 
assuming the form $\exp (-b |t|)$ with the observed slope $b = 5.5$ GeV$^{-2}$ 
\cite{H1,ZEUS}.  The factor 0.9 is included in (\ref{eq:a44}) to allow for 
$\omega$ production.  The $I_L$ and $I_T$ integrations over the gluon 
transverse 
momentum are computed from (\ref{eq:a29}) and (\ref{eq:a30}) as described in 
ref.\ \cite{LMRT}.  We checked the stability of the results to contributions 
from the infrared regions of the $dK^2$ and $d\ell_T^2$ 
integrations.  First we 
varied the infrared cut-off around the value $K_0 = 200$ MeV that we used to 
evaluate (\ref{eq:a27}) and (\ref{eq:a28}).  Second we explored the effect of 
varying $\ell_0^2$ around the value $\ell_0^2 = 1.5$ GeV$^2$ that we used to 
evaluate the integrals of (\ref{eq:a29}) and (\ref{eq:a30}).  Recall that we 
use the linear approximation described in ref.\ \cite{LMRT} to evaluate the 
contribution from the region $\ell_T^2 < \ell_0^2$.  We found only a
weak sensitivity to variation of the choice of $\ell_0^2$.  For
instance reducing $\ell_0^2$ to $1$~GeV$^2$ changes the cross sections
by less than 5\%.  We will 
report on the sensitivity to variation of $K_0$ at the end of the section.

We begin by taking the gluon distribution from the MRS(R2) set of partons 
\cite{MRSR}, which correspond to a QCD coupling which satisfies $\alpha_S 
(M_Z^2) = 0.12$.  The parton set with this QCD coupling, found by 
global analysis of deep inelastic and related data (including recent 
HERA measurements of $F_2$), is favoured by the Fermilab jet data 
with $E_T < 200$ GeV \cite{MRSR}.  We first compare our cross section 
predictions obtained with this gluon with the data.  
Then we use different gluon distributions from several recent sets of partons 
to study the sensitivity of $\gamma^* p \rightarrow \rho p$ to the 
behaviour of the gluon.

Note that we use phenomenological gluon distributions which are 
obtained from global fits to deep inelastic experimental data, rather
than ``ab initio'' distributions calculated from theoretical models.
Thus the gluon distributions that we use already incorporate
absorptive effects.
\begin{figure}[htb]
\centerline{\epsfig{file=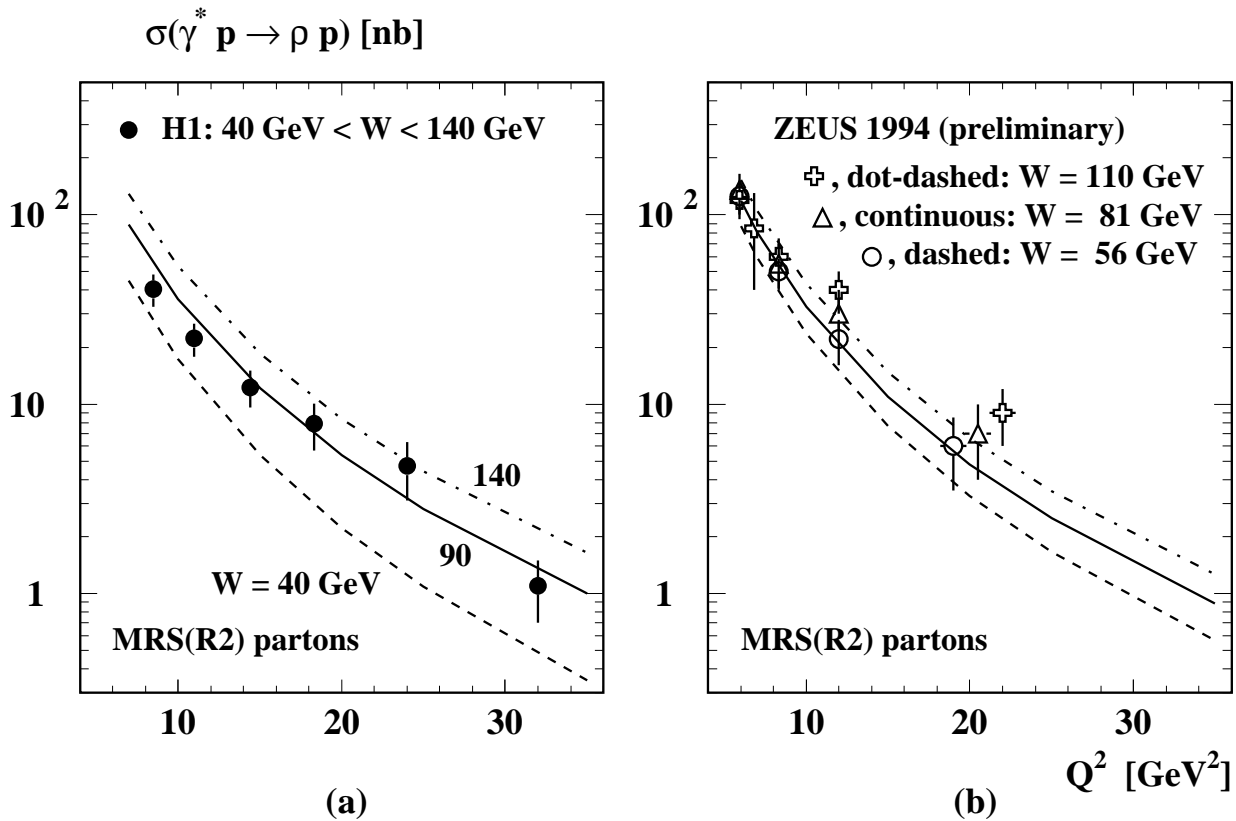,width=18cm}}
\vskip -4mm
\caption[]{ The predicted $Q^2$ dependence of the cross section 
for $\gamma^* p \rightarrow \rho p$ compared with (a) H1 data
\cite{H1} collected over the energy range $40 < W < 140$~GeV and (b)
preliminary ZEUS data \cite{ZEUSW} in energy bins with $\langle W
\rangle = 56,\ 81\ {\rm and}\ 110$~GeV. The QCD curves for the various
values of $W$ are obtained using MRS(R2) partons \cite{MRSR}.}
\label{fig4}
\end{figure}
\begin{figure}[htb]
\centerline{\epsfig{file=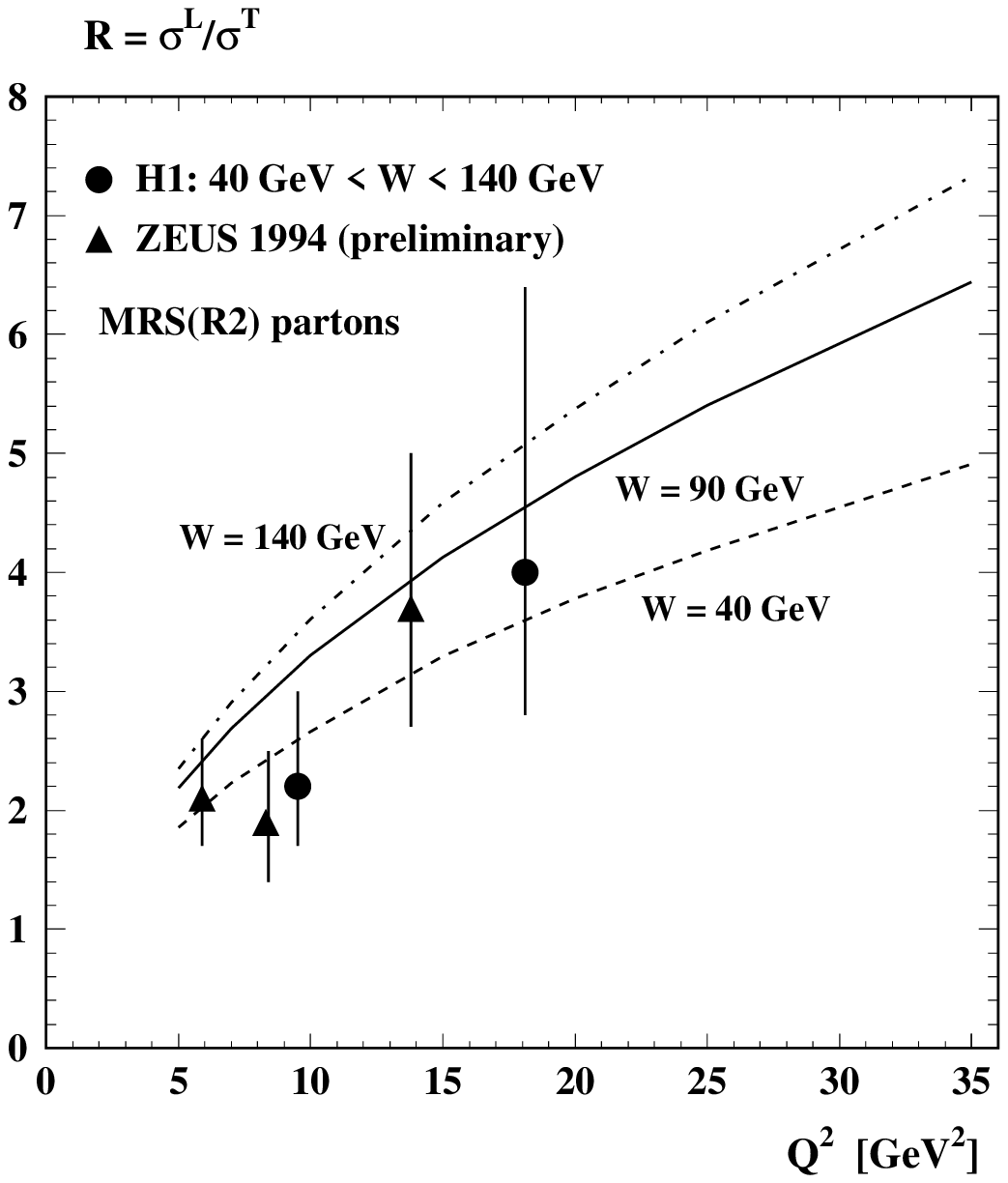,width=11.0cm}}
\vskip -5mm
\caption[]{ The $Q^2$ dependence of the QCD predictions for the 
ratio $\sigma_L/\sigma_T$ of the electroproduction of $\rho$ 
mesons $(\gamma^* p \rightarrow \rho p)$ in longitudinal and 
transverse polarisation states compared with the most recent H1
\cite{H1} and ZEUS \cite{ZEUSW} data. MRS(R2) partons \cite{MRSR} are used.}
\label{fig5}
\end{figure}

There is another crucial ingredient in the calculation of the cross section 
for diffractive open $q\overline{q}$ production.  Virtual gluon corrections to 
the process shown in Fig.~2 are surprisingly important.  The relevant diagrams 
are discussed in ref.\ \cite{LMRT} and lead to $\pi^2$ enhancements of the 
${\cal O} (\alpha_S)$ corrections.  If the contributions are
resummed they lead to an enhancement of the lowest order result by a 
factor $\exp (\alpha_S C_F \pi)$, the so-called $K$ factor enhancement, where 
the colour factor $C_F = 4/3$.  A similar $K$ factor is well known in 
Drell-Yan production, although there the 
contributions come from different virtual 
diagrams \cite{LMRT}.  For the Drell-Yan process the enhancement can 
be as much 
as about a factor of 3.  In our case the $K$ factor can, at present, only be 
estimated.  It proves to be the main uncertainty in the normalization of 
diffractive $q\overline{q}$ production.  The major ambiguity is 
associated with the choice of 
the argument of $\alpha_S$.  We take the scale to be $2K^2$.  Since the $K^2$ 
integrations are dominated by contributions towards the upper limit 
this choice 
is equivalent to a scale $\lapproxeq Q^2/2$.  With this choice we obtain the 
values of the $\gamma^* p \rightarrow \rho p$ cross section shown by 
the curves in Fig.~4, which are in reasonable agreement with the 
measured values.  For our 
choice of scale the average $K$ factor for $\sigma_L$ varies from about 3 to 
3.7 for $Q^2$ going from 25 to 10 GeV$^2$, and is about 20--25\% larger for 
$\sigma_T$ (as in this case somewhat lower $K^2$ values are sampled).  The 
cross section agreement shown in Fig.~4 corresponds to a physically reasonable 
choice of scale, and leads to a sensible range of size of the $K$ factors.  It 
shows that the open $q\overline{q}$ 
duality model for $\rho$ electroproduction is at least consistent with 
observations.  Due to the sensitivity to the choice of scale, clearly 
the agreement cannot be regarded as confirmation of the approach.  
Nevertheless, 
it does imply the existence of a sizeable \lq\lq $\pi^2$" enhancement of the 
Born amplitude, as was also found in the Drell-Yan process.

On the other hand the predictions for the $Q^2$ dependence of the ratio 
$\sigma_L/\sigma_T$ have much less ambiguity.  The calculations are compared 
with the measurements at HERA in Fig.~5.  The agreement with the data shows a 
dramatic \pagebreak
improvement over the QCD expectations which involve convolution with 
the $\rho$ meson wave function.  The small $x$ behaviour of the gluon plays a 
crucial role in masking the $Q^2$ increase anticipated in these earlier 
predictions of the ratio.

\begin{figure}[htb]
\centerline{\epsfig{file=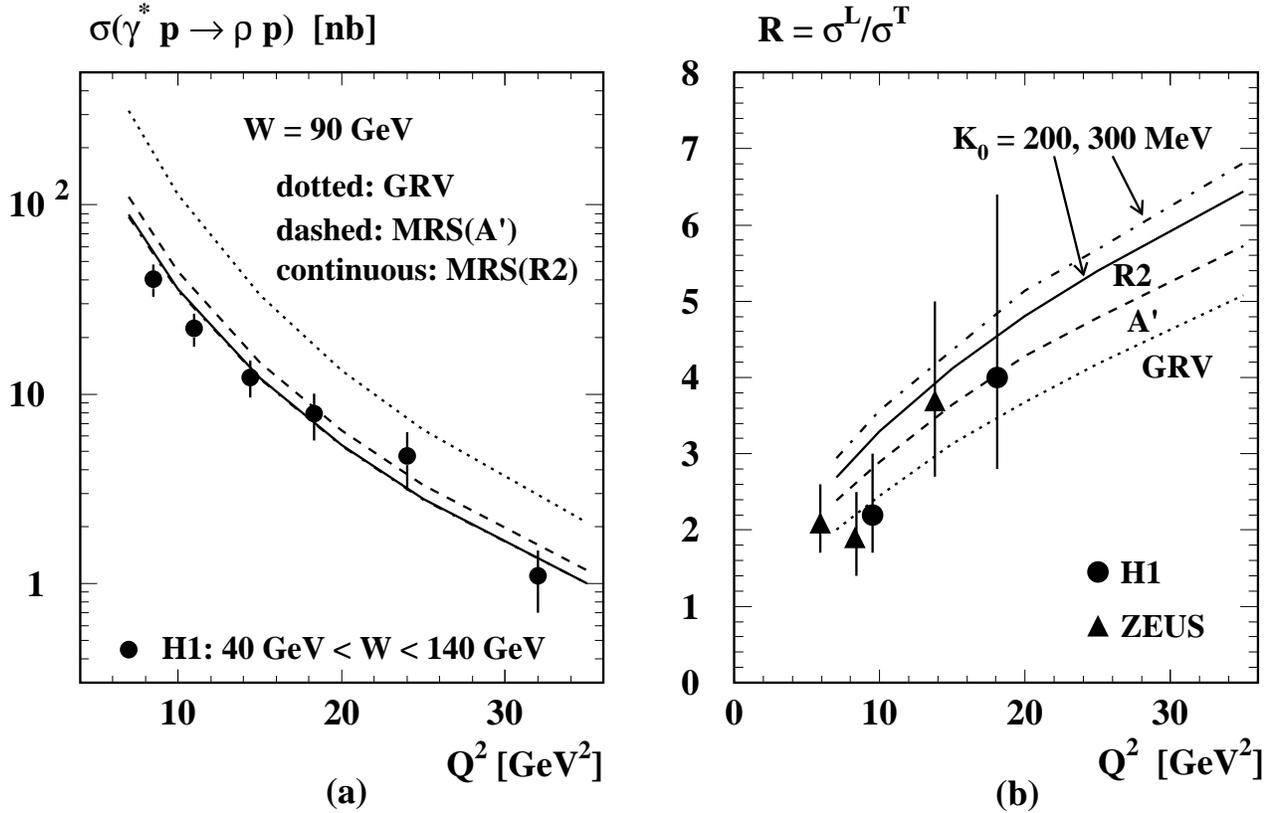,width=18cm}}
\vskip -4mm
\caption[]{The QCD predictions for $W = 90$~GeV based on three recent
sets of partons \cite{MRS,MRSR,GRV} compared with the recent HERA data
\cite{H1,ZEUSW}. We also show the sensitivity of the predictions using
the MRS(R2) partons to the choice of the cut-off $K_0$; the
dot-dashed curves correspond to $K_0 = 300$~MeV whereas all other
curves correspond to $K_0 = 200$~MeV. The dot-dashed curve in (a)
essentially coincides with the continuous curve which demonstrates the
insensitivity of the cross section prediction to the value of $K_0$,
whereas we see that the ratio $\sigma_L/\sigma_T$ of (b) has some
dependence.}
\label{fig6}
\end{figure}

The dependence on the gluon is seen in Fig.~6 which compares the $Q^2$ 
behaviour for $\sigma_L/\sigma_T$ at $W = 90$~GeV for the 
gluon distribution 
of several recent sets of partons (MRS(A$^\prime$) \cite{MRS}, 
GRV \cite{GRV}, MRS(R2) \cite{MRSR}).  We stress that the
normalization of the QCD predictions for the cross section are
dependent on the choice of the mass interval embracing the $\rho$
meson and on the estimate of the $K$ factor enhancement.  On the other
hand the ratio $\sigma_L/\sigma_T$ is not so sensitive to these
ambiguities. 
At this stage it is relevant to study the stability of the results to 
variation 
of the infrared cut-off $K_0$.  This we also show in Fig.~6, where we
present QCD predictions based on MRS(R2) partons for two different
choices of $K_0$.  We see that the cross 
section is hardly changed while the ratio $\sigma_L/\sigma_T$ increases a 
little when $K_0$ is increased from 200 to 300 MeV.  Such a 
result is to be anticipated as $\sigma_T$ samples, on average, 
smaller $K^2$ values than $\sigma_L$.  However, we see that the sensitivity of 
the predictions for $\sigma_L/\sigma_T$ to the value of $K_0$ is sufficiently 
weak so that measurements of the ratio can give a reliable probe of 
the gluon. \\

\noindent {\large \bf 7.  Conclusions}

We have shown that the diffractive electroproduction of $\rho$ mesons at high 
$Q^2$ can be described by perturbative QCD.  Indeed, since $\rho$ 
production in both longitudinally and transversely polarised states 
is being measured at HERA 
with better and better precision, the process $\gamma^* p \rightarrow \rho p$ 
can serve as an excellent testing ground for QCD.  Moreover, we have 
shown that 
it also provides a sensitive probe of the small $x$ behaviour of the gluon 
distribution.

The validity of perturbative QCD is ensured by the large value of $Q^2$.  This 
is already suggested by several features of the existing 
data \cite{H1,ZEUS,NMC}.  
However, the measurements of the ratio $\sigma_L/\sigma_T$ do not support the 
behaviour,
\be
\frac{\sigma_L}{\sigma_T} \; \sim \; \frac{Q^2}{2m_\rho^2},
\label{eq:a45}
\ee
predicted from QCD by convoluting $\gamma^* \rightarrow q\overline{q}$ 
diffractive production with our knowledge of the $\rho$ meson wave function.  
The main problem is that the predictions for $\sigma_T$ are too small and fall 
off too quickly with increasing $Q^2$.  We showed that a non-perturbative 
approach to $\sigma_T$ does not resolve the conflict with the data.  Rather 
we argued that on account of the low mass of the $\rho$ meson the convolution 
with the wave function should be omitted.  The $u\overline{u}$ 
or $d\overline{d}$ 
pairs produced in the $\rho$ mass 
region have, because of phase space restrictions, 
little alternative but to hadronize as $2\pi$ states.  Thus a more appropriate 
approach to $\rho$ electroproduction is to apply the parton-hadron duality 
hypothesis to open $u\overline{u}$ and $d\overline{d}$ production.  Indeed we 
found that this model gives a good description of all of the features observed 
for diffractive $\rho$ electroproduction at HERA, including in particular the 
$Q^2$ behaviour of $\sigma_L/\sigma_T$.  To gain insight into expectations of 
the model, we first made a simple estimate based on assuming a
constant anomalous dimension~$\gamma$.  We found 
\be
\frac{\sigma_L}{\sigma_T} \; = \; \frac{Q^2}{M^2} \: 
\left ( \frac{\gamma}{\gamma + 1} \right )^2,
\label{eq:a46}
\ee
where $M$ is the invariant mass of the $q\overline{q}$ pair 
and $\gamma$ is the 
effective anomalous dimension of the gluon defined by $xg (x, K^2) \sim 
(K^2)^\gamma$, where the typical $K^2$ sampled is $K^2 \lapproxeq Q^2/4$ 
(approximated to be the same for both $\sigma_L$ and $\sigma_T$).  
The decrease of $\gamma$ with increasing $Q^2$ masks the strong growth
shown in (\ref{eq:a45}).  Of course result (\ref{eq:a46}) is greatly 
oversimplified but it gives a good idea of the crucial role 
played by the gluon distribution.  In 
Figs.~4--6 we showed the results of the full calculation.  The computation is 
based on a measured gluon distribution and so automatically 
allows for the appropriate $K^2$ and $x$ dependence of $\gamma$.  The figures 
compare the detailed predictions of the model with the measurements of 
diffractive $\rho$ 
electroproduction at HERA.  The main uncertainty is in the
normalization of the 
cross section.  One source is in the choice of the width of the $\Delta M^2$ 
interval over 
which to apply the duality hypothesis.  The second is associated with the 
$K$ factor enhancement which arises from virtual gluon corrections to open 
$q\overline{q}$ production.  The normalization is sensitive to the 
choice of scale 
used as the argument of $\alpha_S$ in the calculation of the $K$ factor.  The 
data show evidence for a $K$ factor of about 3--4, comparable in size to the 
$K$ factor enhancement established for Drell-Yan production.

The QCD model prediction of the ratio $\sigma_L/\sigma_T$ is essentially free 
of the above ambiguities.  Fig.~6 shows that precise measurements of the ratio 
for $\rho$ electroproduction at different values of $Q^2$, and 
the $\gamma^* p$ 
c.m.\ energy $W$, will provide a valuable probe of the behaviour of the gluon 
distribution $g (x, K^2)$ in the kinematic domain $x \approx Q^2/W^2$ and 
$K^2 \lapproxeq Q^2/4$. 

\newpage

\noindent {\large \bf Acknowledgements}

We thank E.M.\ Levin and R.G.\ Roberts for useful discussions and 
H.~Abramowicz, A.~Caldwell, B.~Clerbaux and A.~De Roeck for
information about the HERA data.  MGR thanks the 
Royal Society for a Fellowship grant and TT thanks the UK Particle Physics and 
Astronomy Research Council for support.  This research was also supported in 
part (MGR) by the Russian Fund of Fundamental Research 96 02 17994.


\begin{thebibliography}{xx}
\bibitem{H1} H1 collaboration:  S.\ Aid {\it et al.}, {\em Nucl.\ Phys.} 
{\bf B468} (1996) 3.

\bibitem{ZEUS} ZEUS collaboration:  M.\ Derrick {\it et al.}, 
{\em Phys.\ Lett.} 
{\bf B356} (1995) 601.

\bibitem{ZEUSW} ZEUS collaboration, submitted paper pa02-028
to the {\em XXVIII.~International Conference on High Energy Physics, 
Warsaw, July 1996}.

\bibitem{NMC} NMCollaboration:  M.\ Arneodo {\it et al.}, {\em Nucl.\ Phys.} 
{\bf B429} (1994) 503.

\bibitem{F2} H1 collaboration:  S.\ Aid {\it et al.}, {\em Nucl.\ Phys.} 
{\bf B470} (1996) 3; \\
ZEUS collaboration:  M.\ Derrick {\it et al.}, {\em Z.\ Phys.} {\bf C69} 
(1996) 607;\\ 
DESY-96-076 and hep-ex/9607002.

\bibitem{MRS} A.D.\ Martin, R.G.\ Roberts and 
W.J.\ Stirling, {\em Phys.\ Lett.} 
{\bf B354} (1995) 155.

\bibitem{MRSR} A.D.\ Martin, R.G.\ Roberts and W.J.\ Stirling, Durham preprint 
DTP/96/44,\\ 
RAL-TR-96-037, hep-ph/9606345, {\em Phys.\ Lett.} {\bf B} (in press).

\bibitem{RY} M.G.\ Ryskin, {\em Z.\ Phys.} {\bf C57} (1993) 89.

\bibitem{BFGMS} S.J.\ Brodsky {\it et al.}, {\em Phys.\ Rev.} {\bf D50} (1994) 
3134.

\bibitem{PHD} E.C.\ Poggio, H.R.\ Quinn and S.\ Weinberg, {\em Phys.\ Rev.} 
{\bf D13} (1976) 1958; \\
Yu.\ L.\ Azimov {\it et al.}, {\em Z.\ Phys.} {\bf C27} (1985) 
65; ibid.~{\bf C31} (1986) 213; \\
Yu.\ L.\ Dokshitzer, V.A.\ Khoze and S.I.\ Troyan, {\em Z.\ Phys.} {\bf C55} 
(1992) 107.

\bibitem{IV} D.Yu.\ Ivanov, {\em Phys.\ Rev.} {\bf D53} (1996) 3564; \\
I.F.\ Ginzburg, D.Yu.\ Ivanov and V.G.\ Serbo, Novosibirsk preprint 
IM-TP-208, Aug.~1995 and hep-ph/9508309;\\
I.F.\ Ginzburg and D.Yu.\ Ivanov, hep-ph/9604437, 
{\em Phys. Rev.} {\bf D} (in press).

\bibitem{CZ} V.L.\ Chernyak and I.R.\ Zhitnitsky, 
{\em Phys.\ Rep.} {\bf 112} 
(1984) 173.

\bibitem{BB} P.\ Ball and V.\ Braun, {\em Phys.\ Rev.}~{\bf D54}
(1996) 2182.

\bibitem{GFS} V.N.\ Gribov, {\em Sov.\ Phys.} JETP {\bf 30} (1970) 709; \\
L.L.\ Frankfurt and M.I.\ Strikman, {\em Phys.\ Lett.} {\bf B65} (1976) 51.

\bibitem{LMRT} E.M.\ Levin, A.D.\ Martin, M.G.\ Ryskin and T.\ Teubner, 
Durham preprint DTP/96/50, hep-ph/9606443, {\em Z.~Phys.} {\bf C} (in press).

\bibitem{NN} N.N.\ Nikolaev and B.G.\ Zakharov, 
{\em Z.\ Phys.} {\bf C49} (1991) 607;\\ 
{\em Phys.\ Lett.} {\bf B260} (1991) 414; 
{\em Z.\ Phys.} {\bf C53} (1992) 
331; \\
J.\ Bartels, H.\ Lotter and M.\ W\"{u}sthoff, {\em Phys.\ Lett.} {\bf B379} 
(1996) 239;\\
E. Gotsman, E.M. Levin and U. Maor, CBPF-NF-021/96, TAUP-2338-96 and 
hep-ph/9606280.

\bibitem{GRV} M.\ Gl\"{u}ck, E.\ Reya and A.\ Vogt, 
{\em Z.\ Phys.} {\bf C67} (1995) 433.

\end{thebibliography}
\end{document}